\documentclass[english]{article}
\usepackage[T1]{fontenc}
\usepackage[latin9]{inputenc}
\usepackage{geometry}
\usepackage{array}
\usepackage{float}
\usepackage{textcomp}
\usepackage{amstext}
\usepackage{graphicx}
\usepackage{rotfloat}

\makeatletter

\newcommand{\lyxmathsym}[1]{\ifmmode\begingroup\def\b@ld{bold}
  \text{\ifx\math@version\b@ld\bfseries\fi#1}\endgroup\else#1\fi}

\providecommand{\tabularnewline}{\\}

\newcommand{\lyxaddress}[1]{
	\par {\raggedright #1
	\vspace{1.4em}
	\noindent\par}
}

\usepackage{makecell}

\makeatother

\usepackage{babel}
\begin{document}
\title{Electropolishing of single crystal and polycrystalline aluminum to
achieve high optical and mechanical surfaces }
\author{A. Brusov, G. Orr\thanks{gilad.orr@ariel.ac.il}, M. Azulay and G.
Golan}
\maketitle

\lyxaddress{Ariel University, Science Park, Ariel 40700, Israel}
\begin{abstract}
Electropolishing has found wide application as the final surface treatment
of metal products in mechanical engineering and instrumentation, medicine
and reflective concentrators for PV cells. It was found that electropolishing
(EP) not only reduces the surface roughness and changes its appearance,
but improves many operational characteristics as well, such as corrosion
resistance, endurance, tensile strength, and many others, and also
changes the physicochemical properties, for example, reflectivity,
electromagnetic permeability and electronic emission of some ferro-magnetic
metals. This fact greatly expands the possibility of using this method
in various fields of science and technology. This work is part of
a study, examining electro polishing for reducing the crystalline
defects adjacent to the surface, thus improving its physical properties,
contributing to higher optical efficiency, when used in PV generation
and storage devices. Five compositions were examined using different
temperature and current density parameters. The polished samples were
evaluated using reflectance spectrometry. The solution which was composed
of Phosphoric acid - 85\%, Acetic acid - 10\%, Nitric acid - 5\% was
found to provide the best results. Another result obtained was that
reflectance increased as the current density increased up to 25 A/dm2.
Further increasing the current density resulted in deterioration of
the surface and reduced reflectance. It was shown that careful lapping
and polishing followed by electropolishing using the suggested solution
may consist of an adequate treatment for preparing reflective concentrators
for PV cells.
\end{abstract}

\section{Introduction}

Electropolishing (EP) is a process of leveling a metal surface by
anodic dissolution and is used to improve the surface quality of metals
in addition to mechanical grinding and polishing. EP is a specific
type of electrolysis that involves a direct electric current passing
through an electrolyte in an electrolytic cell \cite{landolt1987fundamental,han2019fundamental}.
During machining (cutting, grinding) of a metal, the material structure
adjacent to the surface is deformed. In this layer there is a distortion
of the crystal structure, concentration of stresses, inclusion of
foreign materials, defects, which adversely affect the properties
of the material. In thin samples, where the damaged layers contribute
more to the physical properties of the material compared to the undamaged
volume, the effect is pronounced. It is impossible to eliminate this
layer mechanically. Etching is also not acceptable, since it is accompanied
by uneven dissolution of the metal and often leads to its hydrogenation.
EP is the most effective way to remove the thickness-regulated damaged
metal layer and the formation of a new surface layer, devoid of these
disadvantages. EP has a number of advantages compared to mechanical
polishing: short process times, low-cost equipment and consumables,
versatility, and the ability to machine small parts and parts of complex
shapes. EP has worked well for soft metals such as aluminum \cite{vander1999metallography}.
In addition to a well-polished surface, it enhances the materials
characteristics - increases endurance, long-term strength, fatigue
resistance, elastic limit, electromagnetic properties, corrosion resistance
and reduced friction coefficient \cite{inman2013electropolishing,wagner1954contribution,datta2000fundamental,muthoka2015controlled}.
The study of the optical characteristics of soft metal samples puts
forward stringent requirements for the preparation of their surface.
EP has been demonstrated to be the preferred and successful method
to obtain the required optical surface \cite{golovashkin1969optical,shanks2016optics}.
Successful removal of the damaged surface layer is associated with
improvement of the optical, electromagnetic, and some other physical
properties of the material. The classical theories of the EP process
is based on the forming of a viscous liquid layer. In the process
of EP, the transition of a metal into a solution occurs under conditions
of partial passivity, which is associated with the formation of a
viscous liquid layer at the liquid-metal interface which results from
the reaction between the metal and the electrolyte. This high resistance
layer provides non uniform resistance paths with the ridges created
during the mechanical preparation providing lower resistance resulting
in higher currents through them. The higher currents lead to faster
dissolution of the ridges and smoothing of rough surface, which gradually
acquires a wavy relief. Hryniewicz \cite{rokicki2012enhanced} extending
Hoar's work \cite{hoar1950electropolishing} concludes that the oxide
film created on the anode surface supplemented by oxygen introduced
by water decomposition is the only factor in the electropolishing
process. According to this model the metal is dissolved by tunneling
the metal ions via dislocation sites and vacancies of the oxide. As
the thickness of the oxide on the ridges is lower, they dissolve faster.
The effect is a decrease in scattered light and an increase in specular
reflection of light, which manifests itself in the appearance of a
shine on the metal surface. The electropolishing method is used to
prepare samples for studying the microstructure of metals by electron
microscopy \cite{lyman2012scanning} and transmission electron microscopy
\cite{lyman2012scanning,ayache2010sample}. Revealing grain boundaries
on high-purity aluminum is a challenging at times only to be solved
using the EP method. In practice, the use of this method requires
adjusting process parameters depending on the electrolyte, material,
and taking into account the size and shape of the sample. The EP process
parameters that need to be adjusted include: current density, temperature,
time, electrolyte composition and such additional factors as the method
of fastening parts, the material of the cathode and suspension, the
method of isolating the suspension device, the geometric shape and
ratio of the anode and cathode area, and much more, therefore, for
solving a specific research problem, it is necessary to conduct a
series of experiments finding the optimal process parameters. In this
study we have examined the process for removing the damaged layer,
analyzing the surface layer damage using x-ray and specular reflection. 

\section{Experimental procedures}

\subsection{Materials and methods}

Tests were conducted on four types of samples available from different
sources. The first type with dimensions of 30x10x0.3 mm was obtained
from aluminum tape with a minimum purity of 99.99\%. The second and
third samples consisted of commercially available aluminum for sputtering
targets with a minimum purity of 99.99\%, the second with dimensions
of 10 \texttimes{} 12 mm and a thickness of 2 mm, the third with 20
x 10 mm and a thickness of 3 mm. Samples were prepared using electrical
discharge machining. The fourth sample was cut from a single crystal
of aluminum grown in our laboratory, prepared as a 3 mm thick disk.
Sample dimensions were of no significance and were based on available
material or comfort. In a non traditional manner, samples no. 1 and
4 were cut using a 0.3 mm thick diamond cut-off wheel. The first sample
was not lapped since it was manufactured with semi optical surfaces.
Random scratches were present on its surfaces. The second and third
samples were ground and lapped using abrasive paper according to the
following protocol: \#600, \#1000, \#1200, \# 2500, \#4000 grits,
on a polishing/lapping machine. We followed four cleaning protocols
depending on the fabrication stage and the state of the sample. Using
the first protocol, samples were cleaned by immersing the sample into
acetone for 3 minutes in order to degrease it. This was followed by
washing the sample in distilled water and drying it using warm air.
If deeper cleaning was needed, the second protocol consisted of the
samples immersed for 10 minutes in a 10\% solution of acetone and
water in an ultrasonic cleaner, after which they were washed with
running water, followed by washing the samples in ethanol and distilled
water. The process was completed by drying the samples with warm air.
The third protocol was reserved for samples after the EP process.
Samples were washed in ethanol, followed by distilled water, and dried
in warm air. For film removal we developed a fourth protocol in which
the sample was immersed in a 10\% solution of nitric acid, followed
by washing in cold running water, then drying in warm dry air. The
schematic of the electropolishing cell is shown in Figure \ref{fig:Schematic-of-the}.
The anode is a sample of aluminum, and plates of various materials
(lead, graphite, stainless steel) were used as a cathode. The cathode
and anode were immersed in a container with electrolyte; the container
was placed on a heating plate with a magnetic stirrer. The electrolyte
temperature was measured using a mercury thermometer. Voltage was
applied to the electrodes using a high current regulated power supply.

\section{
\begin{figure}[H]
\protect\centering{}\protect\includegraphics[scale=0.7]{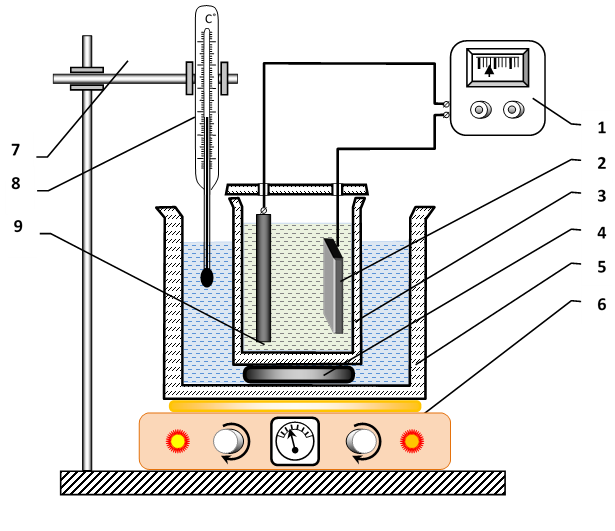}\protect\caption{\label{fig:Schematic-of-the}Schematic of the experimental setup:
1. Power supply; 2. Anode (Aluminum sample); 3. Electrolytic bath;
4. Ceramic stand; 5. Water tank; 6. Magnetic plate; 7. Tripod; 8.
Thermometer; 9. Cathode.}
\protect
\end{figure}
Results and discussion}

\subsection{Electrolyte compositions}

Various electrolytes were used for EP of aluminum. They can be divided
into alkaline and acidic groups. 

\subsubsection*{Alkaline electrolytes }

From the alkaline group, we chose electrolytes with low toxicity -
compositions A and B in Table \ref{tab:Parameters-of-electropolishing}.
The first type of samples were cleaned according to the first protocol,
while the other samples were cleaned more thoroughly using the second
protocol. The cathode consisted of a 12x30x3 mm graphite plate. The
main parameters of the processes are shown in Table 2. 

Some experiment failures were: 1 - points of erosion, deep etching
at point locations; 2 -- An uneven surface finish was obtained. Preferred
etching at the contacts with the sample; 3 -- significant hydrogen
generation at the cathode; 4 - the sample reacts to the electrolyte,
process begins before the voltage is turned on. Some reasons for the
failures are as follows: incorrectly selected electrical parameters,
incorrect estimates of the cathode and anode areas and / or the distance
between them, an unclean surface, poor electrical contact between
the part and the sample holder. Following the above described failures,
the samples were cleaned according to the second deep cleaning protocol,
the design of the sample holder and electrical contact was improved,
the cathode area was increased, and an optimal current density was
selected from the volt-ampere curves for each process.
\begin{table}[H]
\begin{tabular}{ccl>{\centering}p{2cm}>{\centering}p{2cm}>{\centering}p{2cm}>{\centering}p{2cm}}
\hline 
 &  & Electrolyte composition & Temperature $\lyxmathsym{\textdegree}C$ & Current Density range, $A/dm^{2}$ & Voltage range, $V$ & Time, $min$\tabularnewline
\hline 
A &  & Sodium Phosphate 5-10\% &  &  &  & \tabularnewline
 &  & Soda ash 15-30\% & 85-90 & 3-6 & 4-10 & 1-5\tabularnewline
 &  & Water-60-80\% &  &  &  & \tabularnewline
B &  & Sodium phosphate - 320-350 g/l &  &  &  & \tabularnewline
 &  & Sodium carbonate - 230-250 g/l & 80-85 & 2-6 & 5-20 & 10-20\tabularnewline
 &  & Water - 450- 400 g/l &  &  &  & \tabularnewline
\hline 
\end{tabular}
\centering{}\caption{\label{tab:Parameters-of-electropolishing}Parameters of electropolishing
process in alkaline electrolytes}
\end{table}

\begin{figure}[H]
\begin{centering}
\begin{tabular}[t]{lll}
a &  & b\tabularnewline
\includegraphics{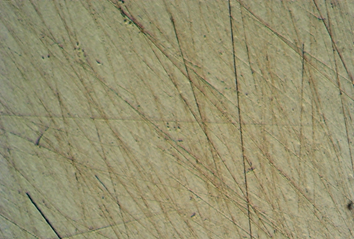} &  & \includegraphics{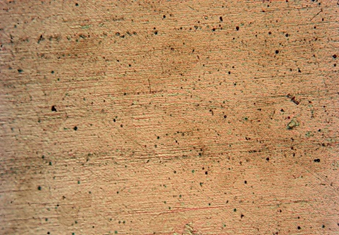}\tabularnewline
\end{tabular} 
\par\end{centering}
\caption{\label{fig:Optical-micrograph-at}Optical micrograph at x 100 magnification
of (a) unpolished aluminum surface; (b) polished aluminum surface
after EP in alkaline electrolyte. }
\end{figure}

Subject to the above conditions, we received a satisfactory result:
defects and scratches left after lapping were removed. Figure \ref{fig:Optical-micrograph-at}
illustrates the microstructure of the aluminum surface before EP (Fig.\ref{fig:Optical-micrograph-at}a)
and after (Fig.\ref{fig:Optical-micrograph-at}b) at x100 magnification.
The polishing quality of samples from pure aluminum and single crystals
is higher than that of aluminum alloy. The advantages of this method
are simplicity, speed and low toxicity. The disadvantages include
insufficient improvement of the surface quality, very high process
speed complicating its control.

\subsubsection*{Acidic electrolytes}

Acidic electrolytes are used more often in world practice than alkaline
ones. Their advantages include: obtaining better reflective surface
properties, increased ability to smooth the microrelief and increase
roughness by 2-3 grades, the possibility of increasing the efficiency
of the process due to the introduction of inhibitors and other additives.
The disadvantage of these electrolytes is the high toxicity of the
reagents. For aluminum and its alloys, electrolytes based on phosphoric,
sulfuric and chromic acids or with chromium anhydride instead of chromic
acid are widely used, as well as electrolytes from a mixture of phosphoric
acid with nitric, acetic or sulfuric acid. The ratio of their components
(mass fraction) fluctuates over a fairly wide range. In experiments,
the influence of various proportions of components on the quality
of electro-polishing was investigated. The compositions of the studied
acidic electrolytes and the parameters of the processes are shown
in Table \ref{tab:Parameters-of-EP}. For example, a mixture of phosphoric
and sulfuric acids with their maximum concentration of compositions
(C and E) leads to the appearance of a dull film consisting of phosphate
salts. This film disappears when the concentration of chromic anhydride
increases from 1\% (electrolyte E) and nitric acid from 5\% (composition
C). An increase in the chromic anhydride content is accompanied by
an increase in the gloss of the metal surface. An increase in chromic
anhydride above 11\% leads to surface etching. The optimum sulfuric
acid content is 26-30\%. In each group of electrolytes, the optimal
ratio of components and process parameters were determined. They are
shown in Table \ref{tab:Parameters-of-the}. 

\begin{table}[H]
\begin{centering}
\begin{tabular}{cl>{\centering}p{2cm}>{\centering}p{2cm}>{\centering}p{2cm}>{\centering}p{2cm}}
\hline 
 & Electrolyte composition & Temperature $\lyxmathsym{\textdegree}C$ & Current Density range, $A/dm^{2}$ & Voltage range, $V$ & Time, $min$\tabularnewline
\hline 
C & Phosphoric acid $H_{2}PO_{4}$ - $45-55\%$ &  &  &  & \tabularnewline
 & Sulfuric acid $H_{2}SO_{4}$ - $30-40\%$ & 100-115 & 30 & 10-30 & 1-2\tabularnewline
 & Nitric acid $HNO_{3}$ - $5-15\%$ &  &  &  & \tabularnewline
\hline 
D & Phosphoric acid $H_{2}PO_{4}$ - $75-85\%$ &  &  &  & \tabularnewline
 & Acetic acid $CH_{3}COOH$ - $10-20\%$ & 100 & 30 & 10-30 & 0.5-1\tabularnewline
 & Nitric acid $HNO_{3}$ - $5\%$ &  &  &  & \tabularnewline
\hline 
E & Phosphoric acid $H_{2}PO_{4}$ - $35-70\%$ &  &  &  & \tabularnewline
 & Sulfuric acid $H_{2}SO_{4}$ - $15-40\%$ & 60-90 & 20-50 & 12-18 & 2-8\tabularnewline
 & Chromic anhydride $CrO_{3}$ - $1-10\%$ &  &  &  & \tabularnewline
 & Water - $10-15\%$ &  &  &  & \tabularnewline
\hline 
\end{tabular}
\par\end{centering}
\caption{\label{tab:Parameters-of-EP}Parameters of EP process in acid electrolytes}
\end{table}

\subsection{Processing parameters}

\subsubsection*{Current density}

\begin{figure}[H]
\centering{}\includegraphics{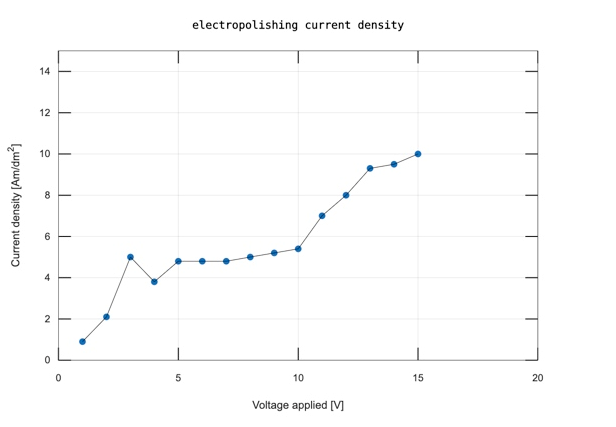}\caption{\label{fig:Electropolishing-current-density}Electropolishing current
density - voltage curve in alkaline phosphate electrolytes for aluminum
sample (type 1)}
\end{figure}

For each one of the electrolytes, the current-voltage relations were
measured. An example of such a curve is illustrated in Figure \ref{fig:Electropolishing-current-density}.
Both alkaline and acids display curves which are similar in form.
The horizontal section of the curve between 4 and 10 volts corresponds
to the process of stabilization of the passivating layer and diffusion
of anions through the passive layer, that is, in this interval, the
polishing process takes place.

\subsubsection*{Temperature}

Different authors recommend different optimal temperatures for the
same electrolyte compositions \cite{tomlinson1958electro,yang2017electropolishing}.
To find out the influence of the temperature factor, samples of the
third type were processed in electrolyte A in the temperature range
80--100°C in increments of 5°C; other process parameters were the
same. The quality of processing was evaluated visually by the presence
of gloss and a comparison of changes in surface topography (the presence
of scratches, and pits) using a microscope at a magnification of x100
times. The amount of metal removed was evaluated by weighing the sample
on an analytical balance. There is a noticeable increase in surface
gloss with the increase of temperature up to 95°C, increasing it to
100°C a decrease in brightness is observed. At the same time the temperature
increase in the above range does not have a noticeable effect on the
metal removal rate or surface relief. To reduce the generation of
hydrogen on the cathode surface, its surface was tripled. 

\subsubsection*{Mixing speed of the solution}

To uniformly heat the electrolyte and improve the separation of gas
bubbles from the surfaces of the electrodes, we applied forced convection
using a magnetic stirrer. The processes was carried out without stirring,
with a low speed, medium, high and maximum. Acceptable results were
obtained when the rotational velocity of the 5 cm magnet was in the
range of 120 to approximately 250 rpm. Without mixing and with a low
mixing speed of the electrolyte, an increase in the number of dot
patterns on the surface is observed. This can be explained by the
fact that the diffusion of matter from the surface of the anode is
difficult and the electrolyte near the sample overheats. A very high
speed leads to displacement of the sample relative to the cathode
and to the deterioration of the contact of the sample holder and the
sample. 

\subsubsection*{The ratio of the surface areas of the anode and cathode and other
additional process characteristics}

EP was carried out with different volumes of electrolyte, with cathodes
of various materials (stainless steel, lead, graphite), and with different
ratios of the areas of the cathode and anode (1: 1; 2: 1; 5: 1). The
best results were obtained by keeping the ratio of the cathode to
the anode at 5: 1, with graphite or lead as the cathode material.
Furthermore, the solution volume should be at least 300 ml for parts
with a surface area of up to 1 $cm^{2}$ and 400 ml for parts with
a surface area of 6 $cm^{2}$, the distance between the anode to the
cathode should be at least 5 cm, and from the sample to the walls
and the bottom of the bath should be at least 3.5 cm.

\begin{sidewaystable}[ph]
\begin{centering}
\begin{tabular}{>{\centering}p{1.5cm}>{\centering}p{1.5cm}>{\centering}p{1.5cm}>{\centering}p{1.5cm}>{\centering}p{1.5cm}>{\centering}p{1.5cm}>{\centering}p{1.5cm}>{\centering}p{1.5cm}l}
\hline 
Elyctrolyte & Sample type & Current density range, $A/dm^{2}$ & Voltage range, V & Temperature, $\text{°}C$ & Time, $min$  & Cathode / anode (ratio) & Mixing speed RPM & Examples\tabularnewline
\hline 
A & 1 & 5 & 10 & 80 & 0.5 & 1:1 & 0 & \textbf{Sample 1,2:} Sodium Phosphate - 10g\tabularnewline
 &  &  &  &  &  &  &  & Soda ash - 22,5g, Water -100ml\tabularnewline
 & 2 & 10 &  & 85 & 1 & 1:2 & 50 & Temperature - $90\text{°}C;$Time 2min;\tabularnewline
 &  &  &  &  &  &  &  & Current density range - 5$A\cdot dm^{-2}$;\tabularnewline
 & 3 & 20 &  & 90 & 2 & 1:5 & 100 & Voltage - 6-9V; Mixing speed - 2-3\tabularnewline
 &  &  &  &  &  &  &  & \tabularnewline
 &  &  &  & 95 & 3 &  & 150 & \textbf{Sample 3:} Temperature - $85\text{°}C$;\tabularnewline
 &  &  &  &  &  &  &  & Time - 1 min; Current density range\tabularnewline
 &  &  &  &  & 4 &  & 200 & $8\,A\cdot dm^{-2}$, other parameters are the same\tabularnewline
\hline 
B & 1 & 5 & 10-30 & 80 & 5 & 1:2 & 100 & \textbf{Sample 1}: Sodium Phosphate - 350 g/l\tabularnewline
 &  &  &  &  &  &  &  & Sodium carbonate - 250 g/l, Water 400 g/l.\tabularnewline
 & 3 & 10 &  & 85 & 10 & 1:5 & 150 & Temperature - $85\text{°C}$, Time 5 min,\tabularnewline
 &  &  &  &  &  &  &  & Current density range $6-9\,A\cdot dm^{-2},$Voltage -6-9V\tabularnewline
 &  & 30 &  &  & 15 &  & 200 & mixing speed 2-3.\tabularnewline
 &  &  &  &  &  &  &  & \textbf{Sample 3}: Temperature - $80\text{°C},$Time - 8 min,\tabularnewline
 &  &  &  &  &  &  &  & Current density range - 15 $A\cdot dm^{-2}$, \tabularnewline
 &  &  &  &  &  &  &  & all the other parameters are as stated above\tabularnewline
\hline 
C & 1 & 25 &  & 90 & 1 & 1:2 & 100 & Electrolyte: Phosphoric acid 50\%,\tabularnewline
 &  &  &  &  &  &  &  & Sulfuric acid - 40\%, Nitric acid - 10\%\tabularnewline
 & 2 & 30 &  & 100 & 2 & 1:5 & 150 & Temperature - $105\text{°C}$, Time 2 min,\tabularnewline
 &  &  &  &  &  &  &  & Current density range - $30\,A\cdot dm^{-2}$,\tabularnewline
 & 3 &  &  & 115 & 3 &  & 200 & Voltage - 12-15 V, Mixing speed - 4\tabularnewline
 &  &  &  &  &  &  &  & \tabularnewline
 &  &  &  &  & 4 &  &  & \textbf{Sample 3}: Time - 1 min, all the other parameters\tabularnewline
 &  &  &  &  &  &  &  & are as stated above\tabularnewline
\hline 
D & 1 & 25 &  & 90 & 05 & 1:2 & 100 & Electrolyte: Phosphoric acid - 85\%,\tabularnewline
 &  &  &  &  &  &  &  & Acetic acid - 10\%, Nitric acid - 5\%,\tabularnewline
 & 2 & 30 &  & 95 & 1 & 1:5 & 150 & Temperature - 100\%, Time - 2 min,\tabularnewline
 &  &  &  &  &  &  &  & Current density range - $30\,A\cdot dm^{-2}$,\tabularnewline
 & 3 & 35 &  & 100 & 2 &  & 200 & Voltage - 12-15V, Mixing speed - 4\tabularnewline
 &  &  &  &  &  &  &  & \tabularnewline
 &  & 40 &  &  &  &  &  & \tabularnewline
\hline 
E & 1 & 20 & 12-15 & 60 & 2 & 1:5 & 100 & Electrolyte: Phosphoric acid 45\%,\tabularnewline
 &  &  &  &  &  &  &  & Sulfuric acid - 35\%, Chromic anhydride - 5\%\tabularnewline
 &  & 30 &  & 70 & 4 &  & 150 & water - 15\%\tabularnewline
 &  &  &  &  &  &  &  & Temperature - $70\text{°C}$, Time 3 min,\tabularnewline
 &  & 40 &  & 80 & 6 &  & 200 & Current density range - $25\,A\cdot dm^{-2}$,\tabularnewline
 &  &  &  &  &  &  &  & Voltage - 12-15 V, Mixing speed - 4\tabularnewline
 &  & 50 &  & 90 & 8 &  & 250 & \tabularnewline
\end{tabular}
\par\end{centering}
\caption{\label{tab:Parameters-of-the}Parameters of the electropolishing process}
\end{sidewaystable}

\subsection{Electrochemical etching}

\begin{figure}[H]
\begin{centering}
\begin{tabular}{lcl}
a &  & b\tabularnewline
\includegraphics[scale=0.6]{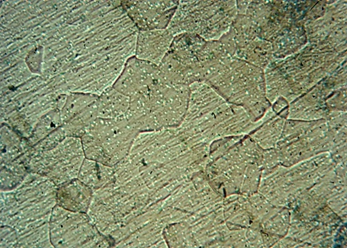} &  & \includegraphics[scale=1.35]{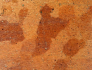}\tabularnewline
\end{tabular}
\par\end{centering}
\caption{\label{fig:Microstructures-of-aluminum}Microstructures of aluminum
after electrolytic etching: a - alkaline electrolyte, b - acidic electrolyte,
magnification $\times100$}
\end{figure}

Our studies have shown the possibility of developing electrolytes
in which it is possible to conduct simultaneous electropolishing and
electro-etching of aluminum to produce a thin oxide layer on it. The
thickness of this layer is minimal, which is necessary to achieve
surface gloss (high reflectivity), but at the same time sufficient
to obtain adsorption staining. The basis of such an electrolyte is
a mixture of sulfuric, phosphoric and acetic acids. For aluminum with
a high degree of frequency, good results are obtained in an aqueous
solution of sodium orthophosphate and sodium carbonate (Figure \ref{fig:Microstructures-of-aluminum}).

\section{results}

\begin{figure}[H]
\begin{centering}
\includegraphics{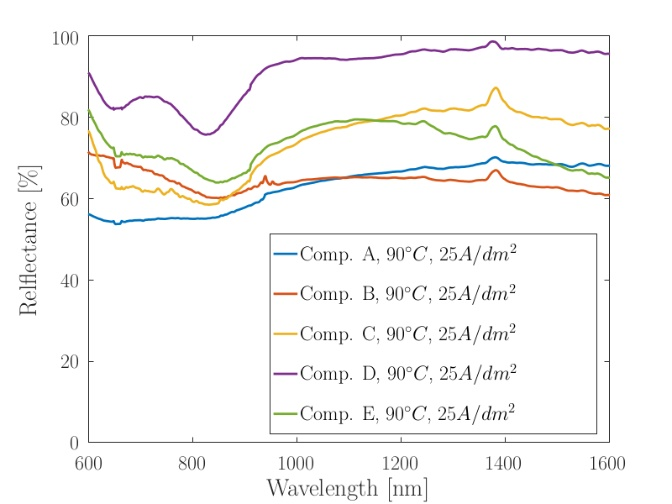}
\par\end{centering}
\caption{\label{fig:A-comparison-of}A comparison of the reflectance of the
various polishing solutions.}
\end{figure}

As one of the primary goals in the current work is reflective concentrators
for PV, it is natural to examine the reflectance in the UV-VIS-IR
range. The measurement system consisted of a reflectance spectrometer
with Deuterium and Tungsten Halogen lamps as sources. Calibration
was conducted using a gold calibration standard thus limiting the
spectral range of the measurement to 600-1600nm. Its strength is the
standards inertness and stability compared to aluminum or silver standards
resulting in improved measurement repeatability. Figure \ref{fig:A-comparison-of}
Compares the reflectance of the various compositions in the tested
range. From the figure it is apparent that in this research, composition
D was superior to the other compositions. The chosen current density
and temperature were found to be optimal in most cases.

\begin{figure}[H]
\begin{centering}
\includegraphics{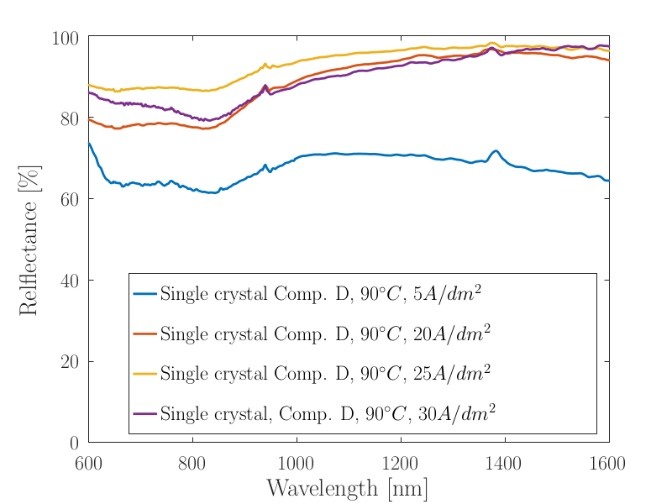}
\par\end{centering}
\caption{\label{fig:Reflectance-measurements-of}Reflectance measurements of
crystals polished using composition D at current densities of $5\,A\cdot dm^{-2}$,
$20\,A\cdot dm^{-2}$, $25\,A\cdot dm^{-2}$, and $30\,A\cdot dm^{-2}$}
\end{figure}

Figure \ref{fig:Reflectance-measurements-of} Illustrates reflectance
measurements of polished crystal samples at several current densities.

It can be seen from the plots that best results were obtained at $25\,A\cdot dm^{-2}$.
Increasing the current from $5\,A\cdot dm^{-2}$ up to $25\,A\cdot dm^{-2}$
displays an increase in reflectance in the formentioned spectral range.
Increasing the current density above $25\,A\cdot dm^{-2}$ displays
reduced reflectance and visual deterioration of the surface. Adelkhani
et al. {[}16{]} reached similar conclusions though demonstrated lower
reflectance at the lower and higher currents. The difference probably
stems from the preliminary processing, i.e. multi stage mechanical
lapping and polishing.

\section{Conclusions}
\begin{itemize}
\item Electropolishing of aluminum is effective for obtaining a high quality
surface and eliminating machining defects. The best result of electropolishing
of an aluminum surface in terms of its quality, as well as the duration
and complexity of the process is provided by a combination of preliminary
mechanical grinding and polishing followed by EP in an acid electrolyte
composition: Phosphoric acid - 85\%, Acetic acid - 10\%, Nitric acid
- 5\% (mode D)
\item We managed to obtain a polished surface of good quality due to the
correct choice of all factors affecting the EP process, such as pretreatment
of the metal surface, orientation of the workpiece in the electrolyzer,
choice of cathode material, distance between electrodes, process time,
solution circulation rate, anode area ratio, and cathode, bath age.
This was confirmed using spectral reflectance analysis.
\item The tested electrolytes, especially acidic ones, have a narrow range
of current density, at which the polishing process takes place. This
limits the ability to control the process: removal of material to
a given depth and obtaining specified roughness parameters.
\item The uniformity of surface treatment is strongly influenced by the
design and material of the suspension, the ratio of the anode to cathode
surface areas and the location of the sample in the cell. In our experiments,
the optimal suspension material is aluminum and the anode-to-cathode
area ratio is 1: 5.
\item The use of electrochemical etching to determine the structure of pure
aluminum gives good results. This is an effective and simple method,
since the process is carried out in the same electrolyzer and with
the same electrolytes as the EP. For high purity aluminum, good results
are obtained in an aqueous solution of sodium orthophosphate and sodium
carbonate.
\end{itemize}
This work was conducted as part of a larger project trying to improve
the aluminum surfaces physical properties in storage devices. As this
article is limited in scope, in a following article we will demonstrate
the improved results obtained in aluminum single crystals prior and
after being processed using the various EP methods discussed in this
article. As shall be demonstrated in future articles, X-ray analysis
indicates that there is a considerable improvement after treating
the surface using the above methods. 

\bibliographystyle{unsrt}
\bibliography{A_Brusov_et_al_2020_Electropolishing_single_crystals_of_aluminum}

\begin{thebibliography}{10}

\bibitem{landolt1987fundamental}
D~Landolt.
\newblock Fundamental aspects of electropolishing.
\newblock {\em Electrochimica Acta}, 32(1):1--11, 1987.

\bibitem{han2019fundamental}
Wei Han and Fengzhou Fang.
\newblock Fundamental aspects and recent developments in electropolishing.
\newblock {\em International Journal of Machine Tools and Manufacture},
  139:1--23, 2019.

\bibitem{vander1999metallography}
George~F Vander~Voort.
\newblock {\em Metallography, principles and practice}.
\newblock ASM international, 1999.

\bibitem{inman2013electropolishing}
M~Inman, EJ~Taylor, A~Lozano-Morales, and L~Zardiackas.
\newblock Electropolishing and throughmask electroetching of nitinol stents and
  other materials in an aqueous electrolyte.
\newblock In {\em Medical Device Materials VI: Proceedings from the Materials
  and Processes for Medical Devices Conference:(MPMD 2011)}, page~31. ASM
  International, 2013.

\bibitem{wagner1954contribution}
Carl Wagner.
\newblock Contribution to the theory of electropolishing.
\newblock {\em Journal of the electrochemical society}, 101(5):225, 1954.

\bibitem{datta2000fundamental}
M~Datta and D~Landolt.
\newblock Fundamental aspects and applications of electrochemical
  microfabrication.
\newblock {\em Electrochimica acta}, 45(15-16):2535--2558, 2000.

\bibitem{muthoka2015controlled}
Boniface~Mutua Muthoka, Alex~Awuor Ogacho, Benard~Odhiambo Aduda, Charles~Opiyo
  Ayieko, Robinson~Juma Musembi, and Pushpendra~K Jain.
\newblock Controlled texturing of aluminum sheet for solar energy applications.
\newblock 2015.

\bibitem{golovashkin1969optical}
AI~Golovashkin, IE~Leksina, GP~Motulevich, and AA~Shubin.
\newblock Optical properties of niobium.
\newblock {\em SOV PHYS JETP}, 29(1):27--34, 1969.

\bibitem{shanks2016optics}
Katie Shanks, Sundaram Senthilarasu, and Tapas~K Mallick.
\newblock Optics for concentrating photovoltaics: Trends, limits and
  opportunities for materials and design.
\newblock {\em Renewable and Sustainable Energy Reviews}, 60:394--407, 2016.

\bibitem{rokicki2012enhanced}
Ryszard Rokicki and Tadeusz Hryniewicz.
\newblock Enhanced oxidation--dissolution theory of electropolishing.
\newblock {\em Transactions of the IMF}, 90(4):188--196, 2012.

\bibitem{hoar1950electropolishing}
TP~Hoar and JAS Mowat.
\newblock The electropolishing of nickel in urea-ammonium chloride melts.
\newblock {\em Transactions of the IMF}, 26(1):7--25, 1950.

\bibitem{lyman2012scanning}
Charles~E Lyman, Dale~E Newbury, Joseph Goldstein, David~B Williams, Alton~D
  Romig~Jr, John Armstrong, Patrick Echlin, Charles Fiori, David~C Joy, Eric
  Lifshin, et~al.
\newblock {\em Scanning electron microscopy, X-ray microanalysis, and
  analytical electron microscopy: a laboratory workbook}.
\newblock Springer Science \& Business Media, 2012.

\bibitem{ayache2010sample}
Jeanne Ayache, Luc Beaunier, Jacqueline Boumendil, Gabrielle Ehret, and
  Dani{\`e}le Laub.
\newblock {\em Sample preparation handbook for transmission electron
  microscopy: techniques}, volume~2.
\newblock Springer Science \& Business Media, 2010.

\bibitem{tomlinson1958electro}
Heather~M Tomlinson.
\newblock An electro-polishing technique for the preparation of metal specimens
  for transmission electron microscopy.
\newblock {\em Philosophical Magazine}, 3(32):867--871, 1958.

\bibitem{yang2017electropolishing}
G~Yang, B~Wang, K~Tawfiq, H~Wei, S~Zhou, and G~Chen.
\newblock Electropolishing of surfaces: theory and applications.
\newblock {\em Surface Engineering}, 33(2):149--166, 2017.

\end{thebibliography}

\end{document}